# INTELLIGENT INTRUSION DETECTION SYSTEM FRAMEWORK USING MOBILE AGENTS

N.Jaisankar[1] and R.Saravanan[2] K. Durai Swamy[3]

[1] School of Computing Sciences, VIT University, Vellore-14.,
njaisankar@vit.ac.in
[2] School of Computing Sciences, VIT University, Vellore-14.,
rsaravanan@vit.ac.in
[3]Dean, K.S.R.C.T., Tiruchengodu, Erode
drkduraiswamy@yahoo.co.in

## ABSTRACT

*An intrusion detection system framework using mobile agents is a layered framework mechanism designed to support heterogeneous network environments to identify intruders at its best. Traditional computer misuse detection techniques can identify known attacks efficiently, but perform very poorly in other cases. Anomaly detection has the potential to detect unknown attacks; however, it is a very challenging task since its aim is to detect unknown attacks without any priori knowledge about specific intrusions. This technology is still at its early stage. The objective of this paper is that the system can detect anomalous user activity. Existing research in this area focuses either on user activity or on program operation but not on both simultaneously. In this paper, an attempt to look at both concurrently is presented. Based on an intrusion detection framework [1], a novel user anomaly detection system has been implemented and conducted several intrusion detection experiments in a simulated environment by analyzing user activity and program operation activities. The proposed framework is a layered framework, which is designed to satisfy the core purpose of IDS, and allows detecting the intrusion as quickly as possible with available data using mobile agents. This framework was mainly designed to provide security for the network using mobile agent mechanisms to add mobility features to monitor the user processes from different computational systems. The experimental results have shown that the system can detect anomalous user activity effectively.*

## 1. INTRODUCTION

In recent years, intrusion and other types of attacks to the computer network systems have become more and more widespread and sophisticated. In addition to intrusion prevention techniques such as authentication, firewall and cryptography used as a first line of defense, intrusion detection is often used as a second line of defense to protect computer networks.

Intrusion detection is defined as "the problem of identifying individuals who are using a computer system without authorization (i.e., `crackers') and those who have legitimate access to the system but are abusing their privileges (i.e., the `insider threat')" [6].

Intrusion detection techniques are traditionally categorized into two methodologies: anomaly detection and misuse detection. Anomaly detection is based on the normal behavior of a subject (e.g., a user or a system); any action that significantly deviates from the normal behavior is considered intrusive. Misuse detection catches intrusions in terms of the characteristics of known attacks or system vulnerabilities; any action that conforms to the pattern of a known attack or vulnerability is considered intrusive.

Mobile agents are programs with persistent identity, which move around a network on their own volition and can communicate with their environment and with other agents. These systems use specialized servers to interpret the agent's behavior and communicate with other servers. Mobile agents may execute on any machine in a network without the necessity of having the agent code pre-installed on every machine the agent could visit. Mobile agents offer several potential advantages when used in ID systems that may overcome limitations that exist in IDS that only employ static, centralized components. The non-monolithic systems based on autonomous mobile agents offer several advantages over monolithic systems [2], such as: easy configuration, extension capacity, efficiency and scalability.

This paper presents a distributed approach to network security using agents to detect the user anomalies. The proposed framework using Mobile Agents is a layered framework, which detects the internal intrusions based on the user and corresponding process profiles.

## 2. MOBILE AGENT PARADIGM

Research in security is most active due to its role in mobile agent system. The issues and requirements of security in mobile agent system have been satisfied by the three basic security principles that a mobile agent system must realize:
- Participants cannot be assumed to trust each other by default.
- Any agent-critical decisions should be made on trusted hosts.
- Unchanging components of the state should be sealed cryptographically.

To implement the mobile agent system under the above security principles, categorized the requirements into four aspects:
- Agent privacy and integrity
- Agent and server authentication
- Authorization and access control

- Metering, charging and payment mechanisms

In this system, the agent can carry three kinds of object, the read-only object or appended only object, they can only be read or appended, any tampering from other agents or server will be detected. As for protecting agents from each other, it is usually done by isolating the execution of agents or providing a facility for agents to authenticate each other (Aglet, Odyssey), providing a secure object space, or using cryptographic techniques to protect objects from tampering. Protecting hosts from malicious agents currently has the most solutions, including various schemes for access control, various authorizations and authentications using digital signatures and other cryptographic techniques, and proof-carrying code.

In most current systems, the host and the Agent Execution environment are assumed to be trustworthy. Those implemented in Java choose to rely on Java's built-in security model and/or provide customized extension of this model.

**Agent migration**

Agent migration introduces agent mobility and portability issues. Portability is generally solved by JVM (Java virtual machine). According to what need to be transferred and who initialized the mobility, agent mobility can be categorized into different degrees, such as remote execution, code on demand, and strong migration.

In Remote Execution, the agent program is transferred before its activation to some remote node, where it runs until its termination. The information transferred includes the agent code plus a set of parameters. In Code on Demand, the destination itself initiates the transfer of the program code. Both Remote Execution and Code on Demand support only code mobility because both schemes transfer agent programs before their activation. The standard agent mobility means migrating agents with not only code but also the state to the destination. The highest degree of mobility is Strong Migration. In this scheme, the underlying system captures the entire agent state (consisting of data and execution state) and transfers it together with the code to the next location. It requires a global model of agent state as well as the transfer syntax for this information.

Moreover, an agent system must provide functions to capture agent state and only few languages allow externalizing the state at such a high level. How to capture and transmit its state, including data as well as execution state is still a problem, especially for low-level thread execution state. The possible alternative is to capture higher-level state in terms of application defined agent data. This strong mechanism might be expensive. In Weak Migration scheme only data state information is transferred. The programmer is responsible for encoding the agent's relevant execution states in program variables. Moreover, the programmer must decide where to continue execution after migration, based on the encoded state information.

Most systems are implemented in java; they extensively used java facilities to realize mobility, such as remote execution, RMI, Java Class loading and object serialization. XML is a new java facility it can be used for mobile migration and has the advantage of realizing fine-grained mobility. The possible mobile agent travelling patterns which we designed while managing a set of managed nodes are itinerary (roaming) model. In the itinerary model the mobile agent generated by mobile agents processor (MAP) has the list of all the managed nodes to be visited. MAP could be characterized as a tool for generating customized mobile agents that are equipped according to the requirements of network manager. By using MAP the functional characteristics of mobile agents, which roam in the network to collect information from managed nodes, can be changed dynamically (i.e. at runtime). Dynamic creation and configuration of mobile agents is achieved using MAP. The design and implementation of mobile agents described in detail is in our previous paper [13].

## 3. RELATED WORKS

A recent survey [10, 11] says that intrusions inside the organizations are growing exponentially. To avoid such intrusions the proposed framework has empowered with mobile agents to detect intrusions in a quick time under the given environment.

In order to detect user anomalies, the normal user activity profiles are created and a similarity score range (upper and lower bound) is assigned to each user's normal pattern set. When in action, the system computes the similarity score of the current activity's patterns, and if this score is not in the similarity score *range*, the activity is considered as an anomaly.

On the user activity level, Bernardes and Moreira[1] has implemented the intrusion detection system using mobile agents. A modular approach is proposed, where independent small agents will monitor the communication paths. The system will have the agents that look for attacks that can be precisely identified by the way they occur i.e., intrusions that follow a well-defined pattern of attack (attack signatures) and these are characteristics of model of improper usage detection (abuse).

On the program activity level, Warrender et al [5] have pointed out that a number of machine-learning approaches such as rule induction, hidden Markov model can be used to learn the concise and generalizable representation of the "self" identity of a system program by relying on the program's run-time system calls. The learned models were shown to be able to accurately detect anomalies caused by exploits on the system programs. Yet, combination of these two for intrusion detection has not been reported.

The combination of user activity and program activity level should produce better results as the system can provide multi-layered information. In this paper, a user anomaly detection system based on the intrusion detection framework [1] has been implemented and conducted several intrusion detection experiments by analyzing both user and program activities.

Initially, the normal user profiles (containing user details and permitted applications) and the program profile (containing the usage of system resource by each user process) have been built. Then mobile agents monitor the current user and program activities from the different hosts and passed to the server, and then load the corresponding normal user profile from the server and compare the current activity patterns with the normal profile to determine the activities are normal or anomaly. Thus, the network intrusion detection can be achieved by providing the security for each individual host. The experimental results have known that the system can detect user anomalies effectively.

This paper is organized in four sections. The introduction of the intrusion detection using mobile agents is presented in the above section 1. In section2, the proposed framework for intrusion detection system is explained. Section 3 describes simulation analysis and results are prescribed. Section 4 gives the conclusion and importance of the paper.

# 4. PROPOSED FRAMEWORK FOR INTRUSION DETECTION SYSTEM

The figure 1 shows the layered approach for proposed framework.

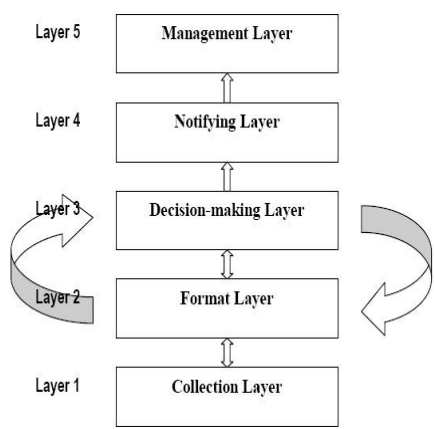

Fig.1: Proposed Framework

The proposed framework (layered model) for the IDS are numbered, starting from the Collection layer (layer 1), and each layer represents a group of specific tasks performed by agents specialized in the functions of that layer. By means of the message exchange mechanism, an agent in a layer activates one or more agents in an upper layer. Based on information collected by Collection Layer, Format layer go into action, by formatting the collected information for the purpose of easy analysis, the Decision-Making layer analyzes and identifies possible intrusions. If these layers suspect an action then they activate Notification Layer, which notifies the network manager and finally the Management Layer manages the profiles according to the incident, and records the complete incidents for future reference.

### 4.1 Collection Layer

This layer allows agents to collect the information related to the users and the processes that the users are using. The following sets of characteristics about user behavior were collected. The set if processes (i.e. number of processes started by user), user login host (i.e. the set of hosts from which user logs on, user session time (the session duration for the user), user activity time (the time of user session starting). At any given point of time hosts are being used by any one of the users. The mobile agent at this layer gathers user and process related information and passes the collected information to the upper layer for further processing. Collection of this information from the host is controlled by agent server and server will activate the agents either on a timely basis or based on an event. The agent in this layer is activated on a timely basis to records user activity and program operation related information from different hosts in the network.

### 4.2 Format Layer

After collecting the information from the layer 1, this layer formats the information and hands it over to the upper layer. This layer is lacking in the earlier framework [1] is introduced in this framework to add pace. Agent server maintains lot of agents and these agents at layer 1 collects information from users. These agents collect raw data from the hosts, which is not in a proper format. This information need to be formatted for easy analysis and decision-making. This layer segregates this information for quicker process. After formatting the data, this layer sends information to the next layer.

### 4.3 Decision – Making Layer

This is the heart of the framework. The agent in this layer collects information from layer 2. Agent server holds the complete information about user profiles and the process profiles of users in a repository associated to it. When the actions are relatively simple, these agents can identify an anomaly or improper use simply by comparing the data obtained with normal profile. To define the normal behavior, we are setting threshold value for normal user and program behavior, based on the more numbers of observations about the behavior, If there is any violation then this layer informs that to the upper layer.

### 4.4 Notifying Layer

The agents of this layer are responsible for notifying the network manager and for activating the agents of layer 5, based on messages received from layer 3. Thus, whenever the Decision-Making Agents identify a level of danger above the acceptable limit or the need to update some new identified pattern, the Notification Agents will be activated. One might think, in principle, that the agents of this layer perform very elementary functions, which would justify (thus eliminating layer 4). However, a decision taken in layer 3 may require several forms of notification, occasioning the construction of a very complex agent, whether its functions are aggregated to this layer or to an upper layer. This would go against the proposed model of small agents performing specific functions in the attempt to minimize the degradation of the environment and reinforce the advantages of an intrusion detection system based on small modules that cooperate with each other.

### 4.5 Management Layer

This layer yet another new addition to the earlier work [1], activate its agents when there is an intrusion. The purpose of this layer is to perform a response when an intrusion occurs. Sometimes it is possible that an intrusion may be detected while it is in progress. As the layer 4 informs the notification to the management layer which activates its agents and tries to stop the intrusion by cutting the privileges to the respect user and blocking the process that caused intrusion.

The figure 2 shows the communication process between the mobile agents.

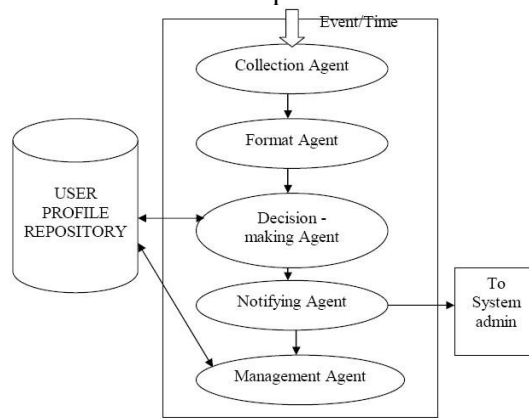

Fig.2: Communication between Mobile Agents

The collection agents are invoked by an event or on a timely basis to collect the information from the host on which user is working. This agent collects all the information about the processes that are being used by the user along with process related information. As it collects the information it submits that information to the upper layer agent, format agent, to segregate the information to reduce analysis time. As it segregates the information and submits that information to the upper layer agent, Decision – making agent, which is associated with the predefined user profiles. This agent compares the collected profile of user at that point of time with the existing profile. If there is any violation of any single parameter then the user is treated as an intruder. If there is no violation then there won't be any response from this agent. When this agent founds an intrusion it informs that to the upper layer agent, notifying agent, who intern informs to the system admin regarding the intrusion. If the intrusion has been modeled already and defined its action after occurring then management agent performs that particular actions.

## 5. SIMULATION AND RESULTS

The proposed framework is being simulated in a simple network environment with one server and minimum of two systems. The software for each type of agent was written in SUN Java JDK Version 1.2 on a Microsoft Windows environment.

The simulation of the proposed framework involves the following
- Building User profile.
- Building process profile for the user.
- Managing user and process profiles.
- Comparing user behavior with the existing profile based on an event or on timely basis.
- Finding intrusion.
- Notifying the intrusion.
- Action to be taken when intrusion occurs.

### 5.1 User Profiles

Building of user profile deals with the essential details like user identification, user authentication and level of the user. This simulation manages the user profiles in three levels. These levels help to take action when an intrusion occurs. This simulation segregates users into three levels, they are: LEVEL 0, LEVEL 1 and LEVEL 2. The user profile of a particular user is shown in figure 3.

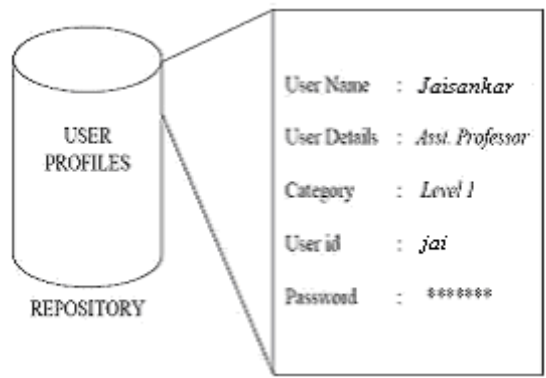

Fig.3: A Simple User-Profile

LEVEL 0 users are usually high priority users in the networking environment like administrators, managers and some other high level authorities. They have unanimous access to all the processes. LEVEL 1 user is the middle level users with middle priority in the organization. Usually most of the users fall under these categories. For this category access to a process depends on their work. Process, which is not related to their work, is not allowed to access. LEVEL 2 users are the least priority users who have very little access to the processes. For example, a temporary guest to the networking environment treated under this category. All the users must fall under any one of these categories.

**5.2 Process Profiles**

All the users in this simulation are associated with information about software applications that have been used during a login session i.e. a process profile, which describes what processes, should be used by the users and some other parameters associated with those processes. This simulation uses some specific processes to test the framework and assigns these processes to the users according to their levels. Each user must contain a process profile with the necessary information about the processes he is using in repository. It is not mandatory that users with same level must have similar process profile. The current simulation uses application–based data source, which is known as process profile in this scenario. The static model is designed to implement the normal behavior of the user and its corresponding process profiles.
The figure 4 shows the process profile of a particular user which contains information about all the application process that user got allocation.

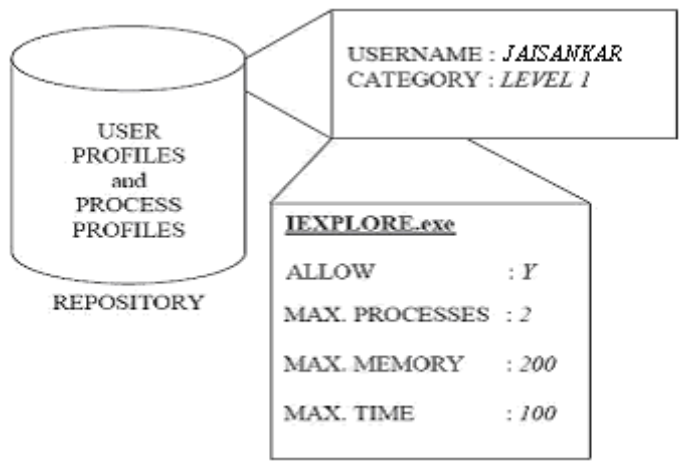

Fig.4: A snapshot of User and Program operation profile

Complete Contents of a process profile for a user is given in the following table 1, which contains information regarding authorized process, the maximum time the user can use the process, the maximum number of times that a process can be used on a particular login session and maximum memory allotted for each process.

| PROCESS NAME | Allow | MAX TIME | MAX PROCESSES | MAX MEMORY | USER ID |
|---|---|---|---|---|---|
| cmd | y | 100 | 10 | 2000 | jai |
| Iexplore | y | 60 | 4 | 2000 | jai |
| wmplayer | N | 50 | 1 | 5000 | jai |
| mspaint | N | 60 | 1 | 200 | jai |
| notepad | y | 100 | 8 | 1000 | jai |

Table 1: Contents of a Process Profile in a Repository

**5.3 Finding Intrusion**

The repository consists of the following information. The user details and the process details for the particular user, i.e. whether that user is authorized to access the particular software applications (process) or not and the usage of the system resource by each process. The collection agents keep track of user activity by recording all related events such as process start, process exit, etc; and on the program activity level, consumption of system resources by user processes are monitored. Intrusions in this simulation are being found by comparing the details of the original existing profile of a user with the data gathered by the agents of that user on a host. If there is any violation, it is treated as an intrusion and activates notifying agents. Here intrusion refers anomaly intrusion detection. In this scenario the right of protecting network starts from the process, by managing these process profiles of users protecting the host, implementing similar things in all the hosts and this way providing security to the total network.

The model of anomaly detection system's profile comparison is shown in figure 5.

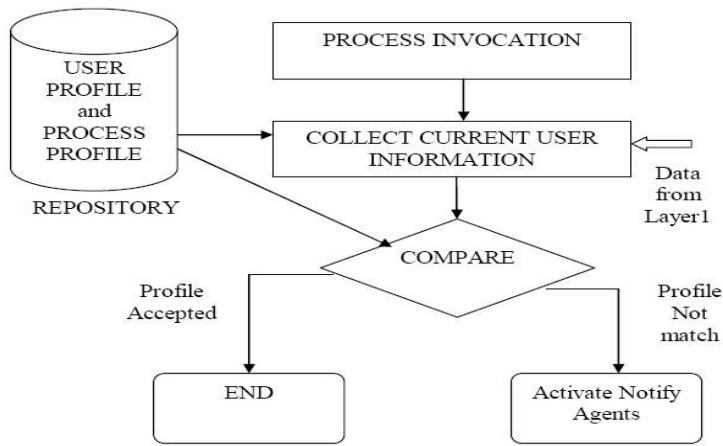

Fig.5: Profile Comparison Procedure

When the user logs on into the particular system, the user id and system details are communicated to the manager in the server side. And color of the process details are indicated into green on screen in simulation analysis, which represents authorized access. If the user accesses the unauthorized software applications (process) which is not the repository, the simulation changed the color green to red, which shows intrusion has occurred. Similarly, the usage of the system resources by the user process exceed the limit, the simulation changed the color green to red. The manager can take the decisions whether the particular user can access the software applications or not by modifying the profiles. The snapshot in this section shows the above scenario.

The overall simulation system structure can be depicted in figure 6.

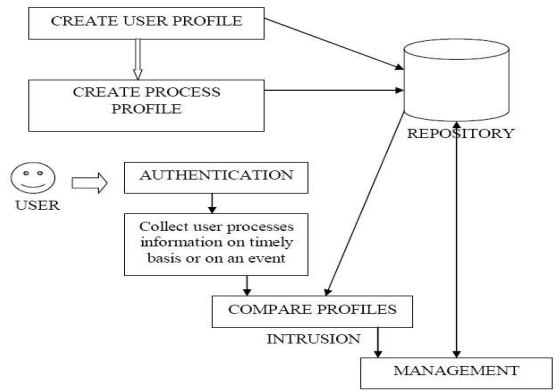

Fig.6: Overall Simulation Structure

We have done the analysis for different time periods such as 30min, 45min, 60min, 80min, 100min and 120min respectively. In specific period, we created number of intrusions and with help of simulation. At all the time periods the simulation showed above 95% ability to detect the intrusion. Analysis has been shown in table 2.

| Time Period (min) | No. of Intrusions created | No. of Intrusions detected | Detection Rate |
|---|---|---|---|
| 30 | 10 | 10 | 100% |
| 45 | 15 | 15 | 100% |
| 60 | 20 | 19 | 95% |
| 80 | 30 | 29 | 96% |
| 100 | 40 | 38 | 95% |
| 120 | 50 | 48 | 96% |

Table 2: Time period versus Detection rate

The following are some of the snapshots of the simulation, which gave most efficient and effective results.

*1. Creation of user profile:*

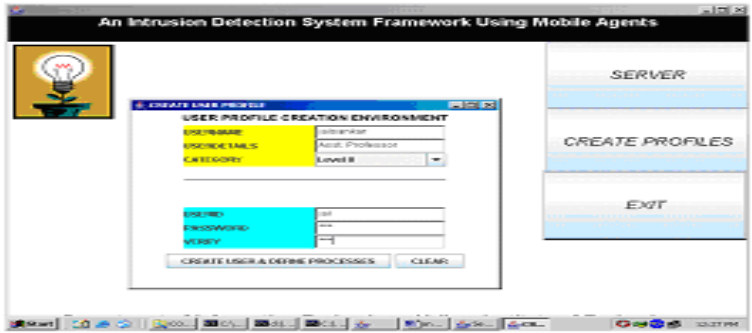

*2Creation of process profile*

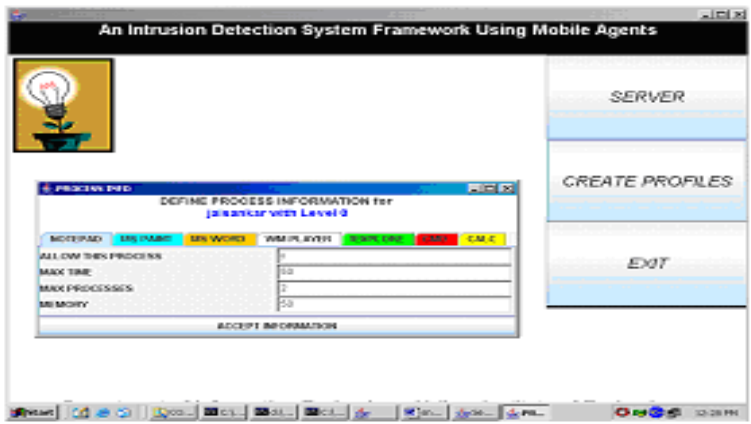

*3. Detection of intrusion when time exceeds for particular process*

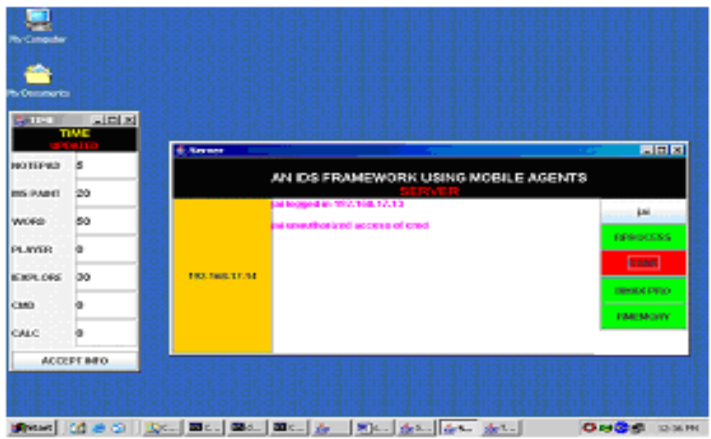

*4. Detection of intrusion for unauthorized for particular process*

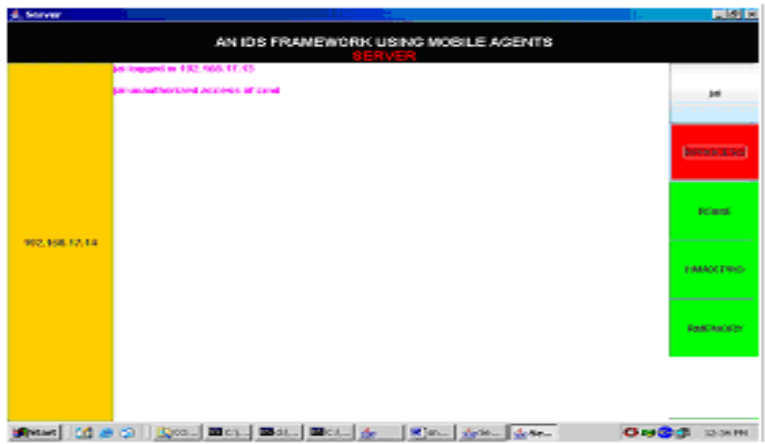

# 6. CONCLUSION

This paper presents a network intrusion system framework using mobile agent, which is able to detect user anomalies in two levels: user activity and program operation. On the user level, the system can detect unauthorized use of programs correctly and on the program level, the excessive use of system resources can be detected which has been shown in section 3.3.

The colors of the process details are indicated into green in simulation analysis, which represents authorized access. If the user accesses the unauthorized software applications (process), which is not the repository, the simulation changed the color green to red, which shows intrusion has occurred. Similarly, the usage of the system resources by the user process exceed the limit, the simulation changed the color green to red. This framework consists of the use of a large number of small mobile agents, which operates independently from the others; however, they all cooperate in monitoring the system, forming complex IDS. In specific period, number of intrusions were created and with help of simulation. At all the time periods, the simulation showed above 95% ability to detect the intrusion. This simulation has been implemented in a simple network, took very less time to detect the intrusion when it occurred or as in progress. This work can be extended in the large network scenario with hundreds of users the performance of the framework. The simplicity and ease of adaptation allows this framework one of the best frameworks in the field of IDS.

**Authors**

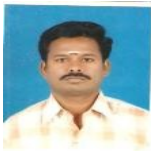

N.Jaisankar is an Associate professor in the School of Computing Sciences at V.I.T. University, Vellore, India. He received his BE (Computer Science and Engineering) from Bharathiar University in 1992. and M.E.(Computer Science and Engineering) from Madurai Kamarajar University in 1999. He has 15+ years of experience in teaching and research. Currently he is perusing research in Network Security. He is a Cisco Certified Network Associate Instructor and SUN certified JAVA instructor. He has published papers in many International and national level conferences and Journals on Network Security and Computer Networks. His research interest includes Network Security, Data Mining, and Computer Networks, Wireless Mobile Ad hoc Network. He has also served in the committees of several international conferences. He is a life member of Indian Society for Technical Education, India and Computer Society of India. He can be reached at njaisankar@vit.ac.in.

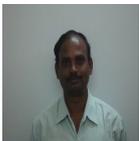

Dr.R.Saravanan is a senior professor in the School of Computing Sciences at V. I. T. University, Vellore, India. He received Ph.D. from Ramanujan Institute for Advanced Study in Mathematics, University of Madras in 1998. He has about sixteen years of teaching and research experience. His areas of specialization include digital image processing, mobile computing and computer networks and security. He is a life-time member of the Computer Society of India and Ramanujan Mathematical Society. He has published fourteen papers in various International and national journals.

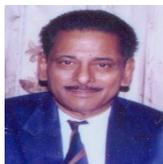

Dr K. Duraiswamy received the B.E. degree in Electrical and Electronics Engineering from PSG College of Technology, Coimbatore in 1965 and the M.Sc. (Engineering) degree from PSG College of Technology, Coimbatore in 1968 and the Ph.D. degree from Anna University in 1986. He has 40+ years of experience in teaching and research. He was the Principal of K.S. Rangasamy College of Technology (KSRCT), Tamil Nadu, and India for 10 years and presently he is serving as the Dean of Academic Affairs in KSRCT. He is interested in Computer Networks, Network Security, Digital image processing, and compiler design. He received 7 years long service gold medal for NCC. He is a life member in ISTE, senior member in IEEE and a member of CSI. He has published over 150 papers in various international, national journals and conferences.